# Observation of the predicted charge collective mode of the $T_c$ = 45 K superconducting phase of $La_2CuO_{4+\delta}$

Y. H. Kim[1*], Y. S. Song[2], and P. H. Hor[2**]

[1]Department of Physics, University of Cincinnati, Cincinnati, Ohio 45221-0011, U.S.A.

[2]Department of Physics and Texas Center for Superconductivity

University of Houston, Houston, Texas 77204-5005, U.S.A.

We report the far-infrared (far-IR) observation of the Goldstone mode at ~ 72 $cm^{-1}$ (~ 9 meV) predicted to exist in the superconducting phase of the transition temperature ($T_c$) at 45 K in the $La_2CuO_4$–based superconductors. Our observation furthers the experimental support for the two-component picture where the localized charge texture, formed at a specific planar hole density ($P_{pl}$), is tied to the HTS at $T_c$ = 15 K, 30 K, and 45 K in a hierarchical fashion at the so-called "magic" doping level at $P_{pl}$ = 1/16, 2/16, and 3/16 respectively.

* kimy@ucmail.uc.edu
**phor@uh.edu

After more than two decades of searching for the mechanism of high temperature superconductivity (HTS) in cuprate superconductors, it is evident that a proper understanding of the inhomogeneous distribution of the doping-induced charge carriers (holes) in the $CuO_2$ planes and their interactions with the antiferromagnetic (AF) spin background is the key to this enigmatic problem. There have been two competing views on the nature of the holes: one suggests that the holes are itinerant and strongly correlated, interacting with some sort of bosonic excitations to form Cooper pairs or simply existing in the resonating valence bond state [1] (one-component picture); the other focuses on the observation that the holes exist in two components, localized and itinerant. Furthermore, the two-component view argues that the localized holes are organized into charge/spin texture [2 – 5]. In the two-component picture this charge/spin texture could exist as a competing order against the superconducting order, essentially preserving the one-component view for HTS [2, 4] or cooperatively interact with the itinerant holes to produce HTS [3, 5].

The necessity of the two-component view was realized first in the early days of HTS when the far-IR reflectivity data was analyzed using the Drude model and the localized hole contribution [6, 7]. However this view was quickly dismissed in favor of the one-component Drude model that could also fit the data by making the scattering rate of the holes frequency dependent. On the other hand, the electrochemical doping study of $La_2CuO_{4+\delta}$ and $La_{2-x}Sr_xCuO_{4+\delta}$ found that when increasing $P_{pl}$ in small steps by fine tuning of the charging potential a protected electronic state of the holes exists at $P_{pl} \sim 0.06$ beyond which a sudden change of the doping efficiency sets in with a chemical potential jump at $P_{pl} \sim 0.06$ [8]. HTS then emerges when $P_{pl}$ exceeds 0.06 in only two distinct superconducting states in $La_{1.985}Sr_{0.015}CuO_{4+\delta}$ system, one at $T_c$ = 15 K and the other at $T_c$ = 30 K [9, 10].

The far-IR reflectivity measurement of the same $La_{1.985}Sr_{0.015}CuO_{4+\delta}$ at $\delta$ = 0.024 ($P_{pl} \sim 0.063$ and $T_c$ = 16 K) [11] revealed that the protected electronic state formed at $P_{pl} \sim 0.06$ accompanies a charge collective mode at $\omega_{G1} \sim 23$ cm$^{-1}$ (~ 3 meV) coexisting with a Drude-like itinerant hole conductivity. This charge collective mode was identified as the Goldstone mode [12] of the broken translational symmetry state of the holes in the $CuO_2$ planes. The finite frequency of the Goldstone mode at $\omega_{G1} \sim 23$ cm$^{-1}$ (~ 3 meV) then indicates that this mode is gapped (massive) and the broken translational symmetry state of the holes is, therefore, insulating [11]. The sum rule analysis suggests that more than 99% of the holes contribute to the Goldstone mode, leaving only less than 1% of the holes itinerant. Therefore, the protected electronic state of the holes at $P_{pl} \sim 0.06$ may be identified as the collective charge texture (CT1) formed at $P_{pl} \sim 0.06$ that is pinned to the underlying $CuO_2$ lattice.

The next naturally arising question is the relevance of the insulating charge texture to the superconductivity. Dumm *et al.* [13] painted the competing order scenario between the static one-dimensional (1D) charge stripes and the superconducting order from their far-IR work on $La_{2-x}Sr_xCuO_4$. Lucarelli *et al.* [14] claimed the far-IR evidence of 1D charge stripes as an essential ingredient for HTS in $La_{2-x}Sr_xCuO_4$ while, through their electrochemical doping and far-IR study of Sr and O co-doped $La_{1.985}Sr_{0.015}CuO_{4+\delta}$, Kim and Hor [11] proposed the two-dimensional (2D) charge textures in the form of square lattice as the ground state on which HTS exists [11, 15]. Therefore, although the far-IR in-plane anisotropy measurements of the detwinned $La_{2-x}Sr_xCuO_4$ single crystals imply that the charge texture is 2D [16], arriving to consensus on the topology of the charge texture is still at the heart of the HTS problem.

Parallel to the broken translational symmetry of the holes by the charge texture in the charge channel, the elastic neutron scattering measurement of the excess oxygen doped $La_2CuO_{4+\delta}$ [17] found that the spin density wave (SDW) ordering takes place at a temperature ($T$) that coincides with the $T_c$ of the system. The inelastic neutron scattering studies of $La_{2-x}Sr_xCuO_{4+\delta}$ system revealed the spin resonance modes at ~ 3 meV (~ 24 $cm^{-1}$) [18] and at ~ 9 meV (~72 $cm^{-1}$) [19] at $T \leq T_c$. These observations led to a possibility that the spin resonances are may be the spin collective modes of the SDW's. Furthermore, since the formation of a charge texture in the AF background of the $CuO_2$ plane is expected to define the spin texture or vice versa, one may make connection between the spin resonance mode at ~ 3 meV (~ 24 $cm^{-1}$) with the Goldstone mode at $\omega_{G1}$ ~ 23 $cm^{-1}$ of CT1.

Therefore the observation of the spin resonance at ~ 9 meV (~ 72 $cm^{-1}$) in $La_{1.82}Sr_{0.18}CuO_4$ ($T_c$ = 37K) [19] is significant because a series of Goldstone modes that are related to the discrete $T_c$'s appear in the charge channel as $P_{pl}$ increases. When $\delta$ in the $La_{1.985}Sr_{0.015}CuO_{4+\delta}$ increased from 0.024 ($P_{pl}$ ~ 0.063) to 0.032 ($P_{pl}$ ~ 0.07), a rapid increase of $T_c$ to 26 K resulted and a new Goldstone mode associated with a new charge texture (CT2) was observed at $\omega_{G2} = 2\omega_{G1}$ ~ 46 $cm^{-1}$ (~ 6 meV) in addition to the mode at $\omega_{G1}$ ~ 23 $cm^{-1}$ (~ 3 meV) [11]. Moreover, the oscillator strength of the $\omega_{G2}$ mode is about twice as large as that of the $\omega_{G1}$ mode, suggesting that CT2 is formed at $P_{pl}$ ~ 0.12. Since $T_c$ = 26 K superconductivity was obtained in a system where CT1 and CT2 are inhomogeneously mixed, it is safe to conclude that CT2 is related to the $T_c$ = 30 K [11].

Thus it appears that each charge texture formed at a specific $P_{pl}$ supports the HTS at a specific $T_c$. This point was verified through the study of a pure Sr-doped $La_{1.93}Sr_{0.07}CuO_4$ ($P_{pl}$ = 0.07) in which the $\omega_{G1}$ mode dominates the $\omega_{G2}$ mode for T < 20 K to result $T_c$ = 16 K superconductivity [15]. This interpretation prompted the prediction that the Goldstone mode at $\omega_{G3} = 3\omega_{G1} \sim 69 - 72$ cm$^{-1}$ (~ 9 meV) of a higher order charge texture (CT3) of $P_{pl} \sim 0.18$, which coincides with the ~ 9 meV neutron resonance mode of SDW, must also exist in the $T_c$ = 45 K superconducting phase in the $La_2CuO_4$–based compound [3].

In this work, we carried out magnetic and far-IR reflectivity measurements of a supercharged polycrystalline $La_2CuO_{4+\delta}$ ($T_c$ = 45 K). We chose the polycrystalline $La_2CuO_{4+\delta}$ because the electrochemical oxidation of the polycrystalline sample is the only way to achieve a dominant $T_c$ = 45 K superconducting phase in a bulk $La_2CuO_4$–based superconductor. It is worthy to note that contrary to the common notion, far-IR measurements of the polycrystalline sample offer an unexpected advantage over single crystals due to its lower reflectivity in the far-IR region. As it has been demonstrated, the near 100 % reflectivity of single crystal superconducting cuprates in the far-IR region makes it difficult to accurately measure the far-IR reflectivity without losing the information buried in the small fraction of changes in the reflectivity [20]. Furthermore it was proven that electrochemically doped polycrystalline samples are *electronically purer* than single crystalline samples due to the high mobility of the oxygen dopant in the $La_2CuO_4$–based compound [10, 20]. Therefore, even though the polycrystalline sample geometry is not ideal to resolve the far-IR anisotropy, we are able to extract all the *in-plane* spectral features belong to the holes in the $CuO_2$ planes because the far-IR anisotropy information of cuprates is well-documented and the scattering by the individual grains and voids of the polycrystalline sample is negligible in the far-IR limit [21].

Systematic studies of electrochemical intercalation of excess oxygen into $La_2CuO_4$ indicated that long time post-annealing after slow, delicate electrochemical oxidation performed at an elevated *T* was required in order to achieve the thermodynamic equilibrium state [8]. However charging at room *T*, at either a constant potential of 450 mV vs. Ag/AgCl or a constant current density at 50 $\mu V/cm^2$ over the geometrical surface area, creates a metastable super-oxidized $La_2CuO_{4+\delta}$ sample ($\delta > 0.065$) of $T_c$ as high as 45 K [9]. It was necessary to store the sample in liquid $N_2$ the entire time in order to preserve the metastable $T_c = 45$ *K* superconducting phase. At room *T* the time evolution of different phases are relatively slow, as shown in Figure 2 of Ref. [9]. We confirmed that the $T_c = 45$ K superconducting phase remained unchanged after we completed all of the far-IR reflectivity measurements.

As displayed in Figure 1 the supercharged sample prepared using the constant potential method shows the $T_c = 45$ *K* superconductivity coexisting with the $T_c = 30$ *K* and the $T_c = 15$ *K* superconductivities in the zero-field-cooled measurement. The $T_c = 30$ K and $T_c = 15$ K superconducting transitions are absent for cooling of the sample in the magnetic field, *i.e.* the Meissner effect measurement. This implies that the $T_c = 30$ K and $T_c = 15$ K superconducting phases only showed up as the screening current in zero-field-cooled measurement and are not the dominant bulk superconducting phase in the supercharged sample. The $T_c = 30$ K and $T_c = 15$ K exist as minute superconducting phases, most likely, at the grain boundaries of the metastable bulk $T_c = 45$ *K* sample.

Detailed far-IR reflectivities of the sample at various $T$ are shown in Figure 2. The reflectivity increases systematically with decreasing $T$. There is a jump in the reflectivity at $T \sim 40$ K, signaling a possible electronic reorganization of the holes to stabilize the energetically favored charge texture in the metastable sample. The raw reflectivity data is displayed except for frequencies below 14 $cm^{-1}$ where smoothing was unavoidable due to noise. There is a reflectivity minimum at $\omega \sim 18$ $cm^{-1}$ and the Goldstone modes at $\omega_{G1} \sim 23$ $cm^{-1}$ and $\omega_{G2} \sim 43$ $cm^{-1}$ are readily seen. In addition there is a sharp mode at $\omega \sim 72$ $cm^{-1}$ and a weak mode at ~ 110 $cm^{-1}$. The intense structure at ~ 220 $cm^{-1}$ is the well-characterized apical oxygen breathing mode polarized along the c-axis below which the c-axis far-IR reflectivity is low and fairly frequency independent. Therefore, it is safe to conclude that the far-IR features observed for $\omega < 220$ $cm^{-1}$ in this work reflect the charge dynamics in the $CuO_2$ planes.

The real part of the conductivity $\sigma_1(\omega)$ and the real part of the dielectric function $\varepsilon_1(\omega)$ calculated from the reflectivity data via a Kramers-Kronig transformation are displayed in Figure 3. The two modes at 23 $cm^{-1}$ and 43 $cm^{-1}$, indicated as $\omega_{G1}$ and $\omega_{G2}$, are respectively the Goldstone modes of CT1 and CT2 that support $T_c = 15$ K and $T_c = 30$ K superconductivity as discussed. An unprecedented new mode indicated as $\omega_{G3}$ now appears as anti-resonance. The weak mode denoted as $X_1$ also appears as anti-resonance. This $X_1$–mode is the highly in-plane polarized, charge texture–induced infrared-active bending mode of La/Sr atom attached to the apical oxygen [16]. Therefore, considering its energy at ~ 72 $cm^{-1}$ (9 meV) we can only stress that the $\omega_{G3}$ mode must be the Goldstone mode of the predicted CT3 that supports the $T_c = 45$ K superconductivity.

The appearance of the $\omega_{G3}$ and the charge texture-induced $X_1$ modes as anti-resonances demonstrates the fact that a Fano-like interference [22] between an electronic continuum and the Goldstone mode at ~ 72 $cm^{-1}$ and $X_1$ mode exists. As commonly seen in spectra of organic conductors, this charge-phonon mechanism gives a forbidden phonon intense oscillator strength of electronic origin [23]. For instance, in the photo-induced infrared spectra of polydiacetylene that exhibits strong charge-polymer backbone coupling, the symmetric (Raman-active) modes of $C = C$ and $C \equiv C$ appear as anti-resonances in the infrared spectra when they overlap with the bipolaron absorption peak [24]. Since the $\omega_{G3}$ and $X_1$ modes and the electronic background must share the same physical origin in order to exhibit the interference [25], we suggest that the CT3 formed in the $CuO_2$ planes provide the electronic continuum that interferes with the $\omega_{G3}$ and $X_1$ modes.

Calculating $\sigma_1 = \omega \operatorname{Im} \varepsilon(\omega)/4\pi$ with $\varepsilon(\omega) = \varepsilon_\infty + f/\left[\omega_e^2\{1 - \lambda D(\omega)\} - \omega^2 - i\Gamma\omega\right]$ and the phonon response function $D(\omega) = \sum_{i=\omega_{G3}, X_1} \omega_i^2/\left(\omega_i^2 - \omega^2 - i\omega\gamma_i\right)$ from the Rice model [25] gave an excellent fit for $\omega_{G3}$ and $X_1$ modes [26] with a Lorentzian centered at $\omega_e = 80$ $cm^{-1}$ (~ 10 meV) as shown in the inset of Figure 3. Here $f$, $\lambda$, $\Gamma$ and $\gamma_i$ are adjustable parameters. This electronic contribution $\omega_e$ comes from the energy required to promote a hole from the CT3 site to an interstitial site which should be less than a half of the single particle excitation energy gap which is on the order of ~ 40 meV. This energy is consistent with that in the electron gap function observed in a high resolution ARPES experiment [27]. We point out that unlike the case of the $\omega_{G1}$ and $\omega_{G2}$ modes, the conductivity data does not reflect the oscillator strength of the $\omega_{G3}$ mode as it enters in the denominator of the $\varepsilon(\omega)$ as seen in the fit. Details of the fit and its $T$-dependence will be published elsewhere.

It has been known in $La_{2-x}Sr_xCuO_4$ system that there exists so-called "magic" doping level at special values of $P_{pl}$ = 1/16 (~ 0.06), 2/16 (~ 0.12), and 3/16 (~ 0.18) [28]. It was found that each electronic phase at "magic" $P_{pl}$ is the electronic state that supports the superconductivity of $T_c$ = 15 K, 30 K, or 45 K in the order of increasing $P_{pl}$ [29]. However, only the local probes may detect the "magic" $P_{pl}$'s as the electronic inhomogeneity due to the mixing of various charges textures would give rise to apparent intermediate $T_c$'s [20]. In conjunction with the findings of scanning tunneling microscopy studies [30 – 33] one may conjecture the simple geometrical meaning of the "magic" doping $P_{pl}$'s: they belong to the $P_{pl}$ of 2D square lattices that commensurate with the underlying the $CuO_2$ lattice[13]. Therefore, since $P_{pl}$ controls the superconductivity, this observation implies that the coexistence of CT1, CT2, and CT3 together with the charge texture of $P_{pl}$ = 4/16 (CT4) would comprise the peculiar superconducting dome in the electronic phase diagram of the $La_{2-x}Sr_xCuO_{4+\delta}$ system.

The finding of the predicted Goldstone mode ($\omega_{G3}$) in the $La_2CuO_4$–based superconductor completes the sequence of the Goldstone modes $\omega_{G1}$ ~ 23 $cm^{-1}$ (~ 3 meV), $\omega_{G2}$ ~ 46 $cm^{-1}$ (~ 6 meV), and $\omega_{G3}$ ~ 72 $cm^{-1}$ (~ 9 meV) which respectively belong to CT1 ($P_{pl}$ = 1/16), CT2 ($P_{pl}$ = 2/16), and CT3 ($P_{pl}$ = 3/16) that are formed as $P_{pl}$ increases. This intrinsic cuprate physics should also be applicable to entire cuprate family because the same mechanism is in operation behind the HTS observed in all cuprate superconductors. For example, the same "magic" $P_{pl}$ order is also found in the $Y_1Ba_2Cu_3O_{7-\delta}$ system [34] where three intrinsic superconducting phases exist in the charge channel at $P_{pl}$ ~ 0.06 ($T_c$

= 30 K), $P_{pl}$ ~ 0.12 ($T_c$ = 60 K), and $P_{pl}$ ~ 0.18 ($T_c$ = 90 K) [35, 36]. In the spin channel, in addition to the ~ 41 meV neutron spin resonance mode in $Y_1Ba_2Cu_3O_{7-\delta}$ ($T_c$ = 90 K) [37], the neutron resonance mode at ~ 30 meV (240 $cm^{-1}$) in $Y_1Ba_2Cu_3O_{6.5}$ ($T_c$ = 60 K) [38] has been observed. The far-IR study of the same sample found a charge-induced mode at ~ 240 $cm^{-1}$ (~ 30 meV) [39] which is, we suggest, the Goldstone mode of the CT2 of $P_{pl}$ = 2/16. Therefore we are led to predict the presence of the Goldstone modes at ~ 110 $cm^{-1}$ (~ 14 meV) of CT1 and at ~ 330 $cm^{-1}$ (~ 41 meV) for CT3 in the charge channel of $Y_1Ba_2Cu_3O_{7-\delta}$ as summarized in Table 1. In fact there have been ample indications of the presence of the Goldstone modes at ~ 110 $cm^{-1}$ and ~ 330 $cm^{-1}$ in the far-IR and Raman studies of $Y_1Ba_2Cu_3O_{7-\delta}$[40 – 42]. The Goldstone mode at ~ 110 $cm^{-1}$ of $T_c$ = 30 K phase should also induce the neutron resonance mode at ~ 14 meV in the spin channel. In addition, the neutron resonance mode related to CT2 at ~ 6 meV in $La_{2-x}Sr_xCuO_4$ should also be present as listed in Table 1.

In conclusion we report the observation of the predicted Goldstone mode $\omega_{G3}$ at ~ 72 $cm^{-1}$ in a $La_{2-x}Sr_xCuO_{4+\delta}$ containing the dominating superconducting phase of $T_c$ = 45 K. Together with the ~ 9 meV neutron spin resonance mode of $La_{1.82}Sr_{0.18}CuO_4$ ($T_c$ = 37K) [19], our observation of the predicted ~ 9 meV Goldstone mode of CT3 of $P_{pl}$ ~ 0.18 further proves that the charge and spin textures are formed at one of the "magic doping $P_{pl}$'s. The predictions made in Table 1 need to be confirmed in order to establish the connection between the Goldstone mode of charge texture and the spin resonance mode of corresponding SDW texture for T ≤ $T_c$. These observations will not only settle the issues of the two-component picture of the holes but also lay the experimental foundation for the development of the microscopic theory of HTS.

PHH is supported by the State of Texas through the Texas Center for Superconductivity at University of Houston.

**Table 1. The observed and predicted far-IR Goldstone modes and neutron spin resonance modes of superconducting $La_{2-x}Sr_xCuO_{4+\delta}$ and $Y_1Ba_2Cu_3O_{7-\delta}$.**

| | | $P_{pl}$ = 1/16 | $P_{pl}$ = 2/16 | $P_{pl}$ = 3/16 |
|---|---|---|---|---|
| $La_{2-x}Sr_xCuO_{4+\delta}$ | $T_c$ | 15 K | 30 K | 45 K |
| | Far-IR | ~ 23 $cm^{-1}$ ( 3 meV) (Observed[11]) | ~ 46 $cm^{-1}$ (6 meV) (Observed[11]) | ~ 70 $cm^{-1}$ (9 meV) (This work) |
| | Neutron | 3 meV (Observed[18]) | 6 meV (Predicted) | 9 meV (Observed[19]) |
| $Y_1Ba_2Cu_3O_{7-\delta}$ | $T_c$ | 30 K | 60 K | 90 K |
| | Far-IR | ~ 110 $cm^{-1}$ (14meV) (Predicted) | ~ 240 $cm^{-1}$ (30 meV) (Observed[39]) | ~ 330 $cm^{-1}$ (41 meV) (Predicted) |
| | Neutron | 14 meV (Predicted) | 30 meV (Observed[38]) | 41 meV (Observed[37]) |

**Figure Captions**

**Figure 1**. Magnetization of the electrochemically charged $La_2CuO_{4+\delta}$ at 550 mV for 25000 seconds at H = 5 Oe as a function of temperature under zero-field textured cooled (open square) and field-cooled (open red circle) conditions.

**Figure 2**. Far-IR reflectivity of the polycrystalline sample of the electrochemically charged $La_2CuO_{4+\delta}$ . From bottom to top: T = 300 K, 270 K, 250 K, 220 K, 200 K, 180 K, 160 K, 120 K, 100 K, 70 K, 50 K, 49 K, 48 K, 47 K, and 46 K (Gray). T = 45 K, 44 K, 43 K, 42 K, 41K, 40 K, 37 K, and 35 K (Orange). T = 30 K, 25 K, 20 K (Green). T = 15 K, 14 K, 12 K, 10K, 8 K, and 6 K (Black). Notice the jump in the reflectivity between the $T_c$ = 45 K superconducting state and the $T_c$ = 30 K superconducting state (see text for details).

**Figure 3.** Temperature dependence of $\sigma_1(\omega)$ and $\varepsilon_1(\omega)$ calculated from the reflectivity data. The same color code scheme as Figure 2 was used. Upper Panel: $\sigma_1(\omega)$ with the temperature increasing from bottom to top. Note that the Drude-like free hole conductivity that matches the dc conductivity is out of the range of the measurement as the scattering rate is extremely small less than 10 $cm^{-1}$ (see Ref [11] for a detailed discussion). Lower Panel: $\varepsilon_1(\omega)$ with temperature decreasing from top to bottom. The plasma frequency ($\omega_p$) of the free holes is at ~ 18 $cm^{-1}$. Inset: Theoretical fit to $\sigma_1(\omega)$ at $T$ = 6 K by considering $\omega_{G3}$ and $X_1$ modes only with the electronic continuum absorption peaked at 80 $cm^{-1}$ (see text for details).

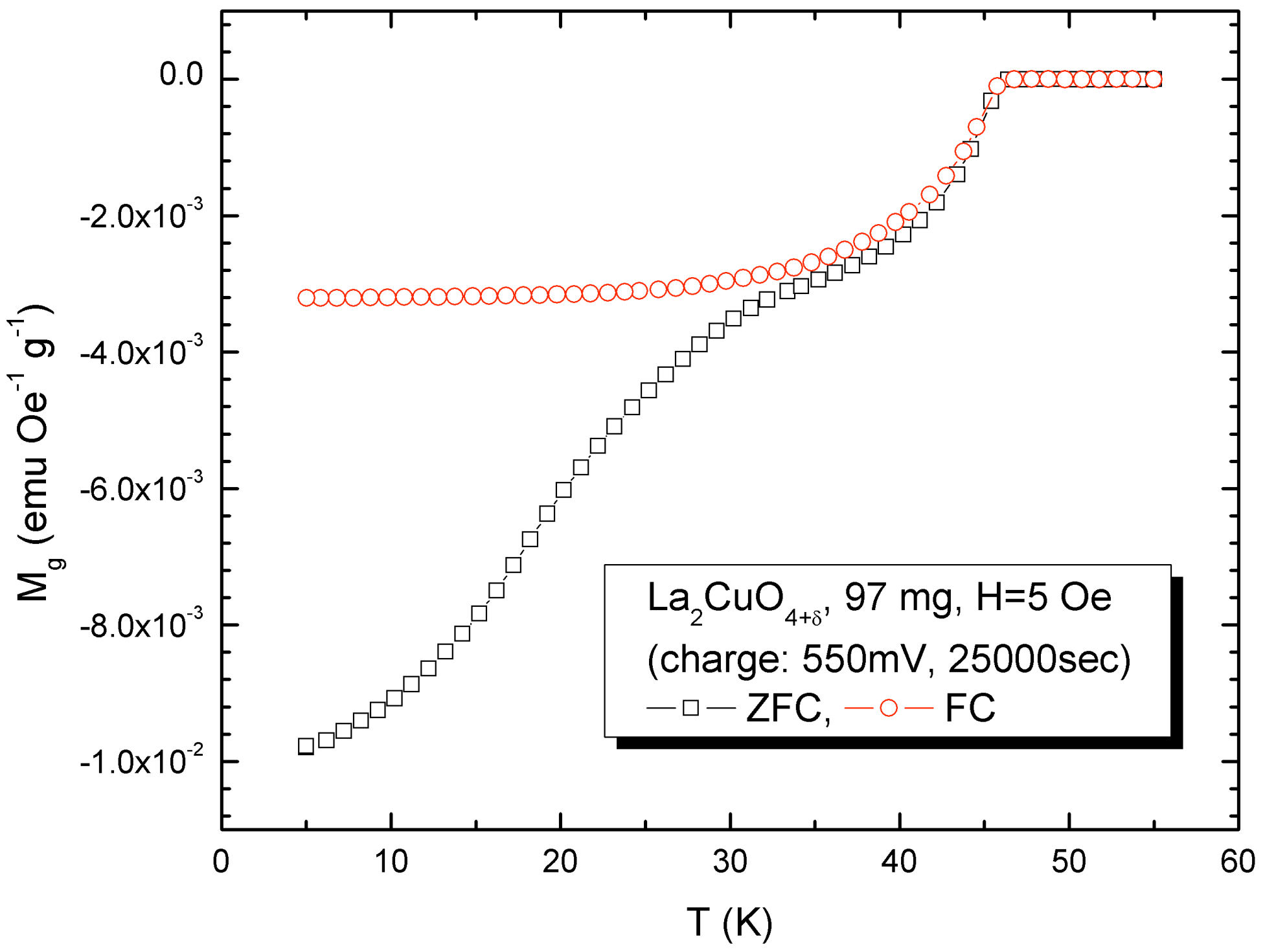

# Figure 1

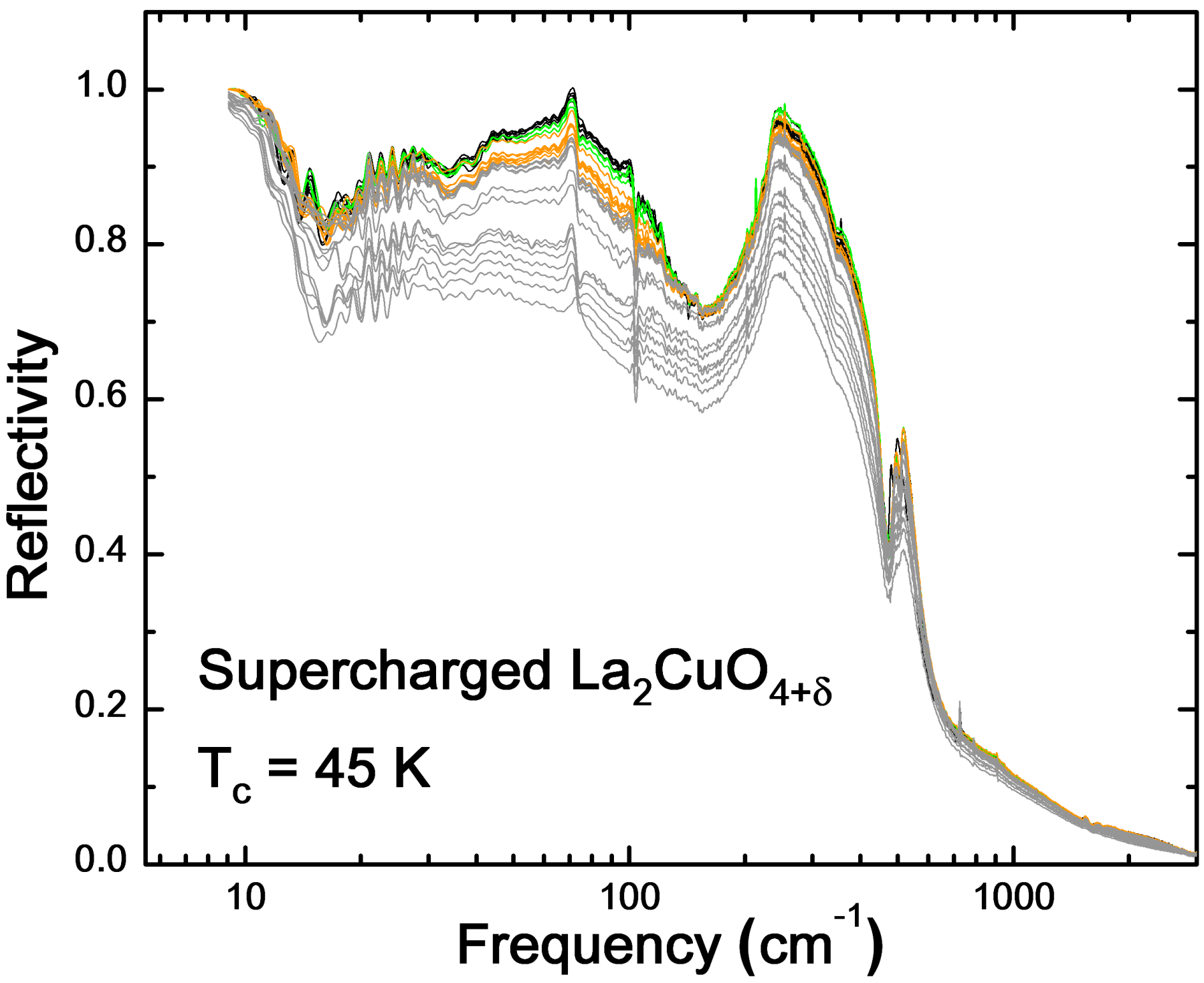

# Figure 2

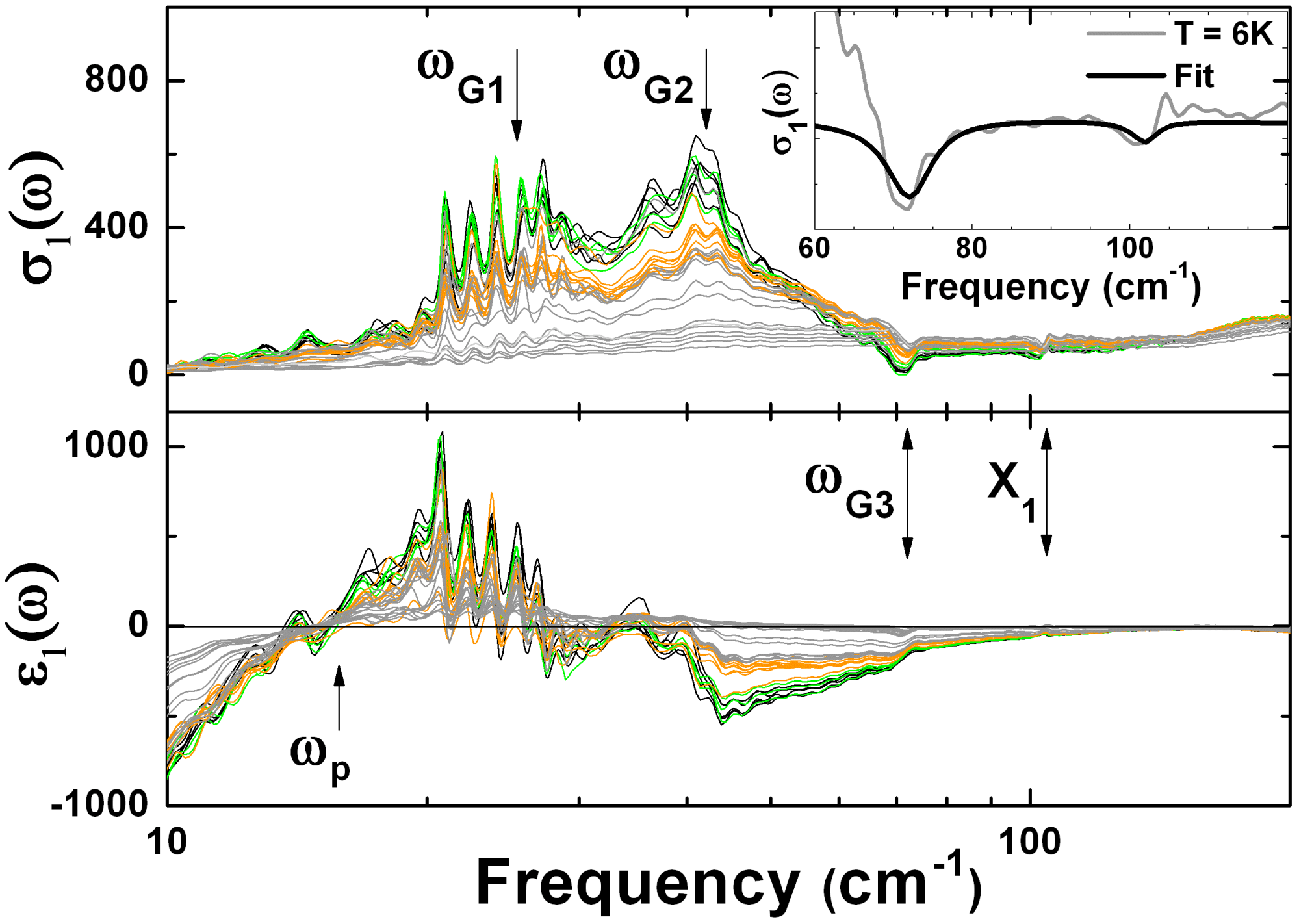

# Figure 3